\documentclass[twocolumn]{revtex4}
\usepackage{amsmath}
\usepackage{amssymb}
\usepackage{graphicx}
\usepackage{dcolumn}
\usepackage{bm}
\usepackage{epstopdf}

\begin{document}

\title{A re-formulization of the transfer matrix method for calculating wave-functions in higher dimensional disordered open systems}
\author{Liang Chen$^{1,2}$}
\email{cq.chenliang@gmail.com}
\author{Cheng Lv$^{1,2}$}
\author{Xunya Jiang$^1$}
\email{xyjiang@mail.sim.ac.cn}
\affiliation{ $^1$State Key Laboratory of Functional Materials for Informatics, Shanghai Institute of Microsystem and Information Technology, CAS, Shanghai 200050, China} \affiliation{ $^2$Graduate School of Chinese Academy of Sciences, Beijing 100049, People's Republic of China\\}

\begin{abstract}
We present a numerically stable re-formulization of the transfer
matrix method (TMM). The iteration form of the traditional TMM is
transformed into solving a set of linear equations. Our method gains
the new ability of calculating accurate wave-functions of higher
dimensional disordered systems. It also shows higher efficiency than
the traditional TMM when treating finite systems.
In contrast to the diagonalization method, our method not only
provides a new route for calculating the wave-function corresponding
to the boundary conditions of open systems in realistic transport
experiments, but also has advantages that the calculating wave
energy/frequency can be tuned continuously and the efficiency is
much higher. Our new method is further used to identify the necklace
state in the two dimensional disordered Anderson model, where it
shows advantage in cooperating the wave-functions with the
continuous transmission spectrum of open systems. The new
formulization is very simple to implement and can be readily
generalized to various systems such as spin-orbit coupling systems
or optical systems.
\end{abstract}
\maketitle



\section{Introduction}
The transfer matrix method (TMM) is a powerful numerical
technical scheme for solving differential equations~\cite{TMMbook}.
In condensed matter physics, it has been successfully used for studying
transport properties of electronic systems for decades~\cite{ROP1,ROP2,ROP3}.
In particular, assisted with the finite size scaling analysis~\cite{FSS},
it is the most frequently used method for analyzing the Anderson
metal-insulator transition in disordered systems~\cite{ROP1,ROP2,ROP3,FSS}.
The methodology is also widely used for electromagnetic waves,
elastic waves and many other systems~\cite{MSF,MoreTMM}.

Despite the success of the TMM, for disordered systems with dimension
higher than one, the method suffers from serious numerical
instability~\cite{ROP1,ROP2,ROP3,MSF}. Since the eigenvalues of the
disordered transfer matrices are of the form $e^{\pm \alpha}$,
as iterating the transfer matrices, the operating vectors rise
exponentially with a set of \emph{different} exponents $\alpha_i$'s.
The vector rising slowest will become inaccurate after $N$ iterations,
provided that $e^{N(\alpha_{min}-\alpha_{max})}<\epsilon$, where
$\alpha_{max}(\alpha_{min})$ is the largest(smallest) exponent and
$\epsilon$ is the accuracy of the computer~\cite{ROP2}.
If only calculating the transmission coefficient $T$,
such a problem can be overcome by applying the so-called
``re-orthogonalization process'', which saves the information
of those vectors contributing largest to $T$(likewise, the Landauer
conductance $g=Te^2/h$ or the localization length)~\cite{ROP1,ROP2,ROP3}.
However, \emph{the numerical instability leads to a rather inflexible
drawback} that the steady-state \emph{wave-function}, which carries fundamental
information for understanding transport phenomena, is inaccessible in
the traditional TMM, since accurate wave-functions demand all vectors
in the TMM iterations to be correct~\cite{MSF1}.
To obtain the accurate wave-function, physicists usually directly diagonalize
the system Hamiltonian with some artificial boundary conditions, such as periodic or
hard-wall boundary conditions~\cite{ROP2,ROP3,Diag}. But, under those boundary conditions,
the system is \emph{closed}(such that the Hamiltonian is Hermitian)~\cite{OpenSystem},
which is essentially different from those \emph{open systems}~\cite{OpenSystem}
treated in the TMM.
An obvious result of the difference is that the transmission and wave-function(responses)
of open systems in the TMM continuously change with the energy/frequency
of incident waves(external excitation), while for closed systems treated in the
diagonalization one can only obtain wave-functions at discrete eigenvalues.
The boundary conditions of the TMM directly correspond to realistic transport experiments,
such as electronic conductance or optical transmission measurements, where the
underlying wave propagation and scattering phenomena are of practical and fundamental
importance~\cite{OpenSystem,OpenSystem1,OpenSystem2}. Nevertheless, currently the
wave-functions according to such boundary conditions cannot be calculated from
either the traditional TMM or the diagonalization.

Recently, the need for the wave-functions of the TMM becomes pressing in
studying transport properties of higher dimensional disordered systems.
Recent studies~\cite{Pendry,Tartakovskii,Bertolotti2005PRL,Sebbah2006PRL,
Bertolotti2006PRE,Ghulinyan2007,Bliokh2008PRL,LiWei2009,Zhang2009PRB,
Chen2011NJP,WangJ} approve that the transmission of the Anderson localized system
is dominated by a kind of \emph{necklace state(NS)} predicted by
Pendry~\cite{Pendry} and Tartakovskii~\cite{Tartakovskii}. The NS is
formed through the coupling between localized states which are nearly degenerate
and spatially near each other.
In transport experiments, it is characterized by the ``chain of
localized states'' from the spatial wave-functions and the continuous
``mini-band'' in the transmission spectra~\cite{Bertolotti2005PRL,Sebbah2006PRL}.
In one dimensional(1D) disordered systems where the numerical
instability does not exist, the TMM has already been intensively used to
study the NS~\cite{Pendry,Bertolotti2005PRL,Bertolotti2006PRE,Ghulinyan2007,
LiWei2009,Chen2011NJP}. It shows the TMM has inherent advantages for studying
those coupled states, since both the ``chain of localized states'' and the
``mini-band'' can be directly observed from the calculated wave-functions and
transmission spectra.
In higher dimensional systems, the NS is also predicted
to exist and even more important role of the NS is predicted~\cite{Pendry,Chen2011NJP,
Mookerjee1993,ShengScience,PendryNature,Vanneste2009PRA}.
Unfortunately, since the wave-functions of higher dimensional TMM have so
far been hard to access, similar TMM studies of the NS in higher dimensional
systems cannot be carried out.
Also, the mini-band, the sign of the NS in spectra, cannot be obtained by
direct diagonalization of the Hamiltonian of closed systems.
In view of such a lack, the application of the TMM could be largely extended
if it could be improved to obtain both accurate wave-functions and continuous
transmission spectra of higher dimensional systems.

In this paper, we present a numerically stable re-formulization of the TMM.
The boundary conditions as well as iterations of the traditional TMM
are embedded in a set of linear equations self-consistently.
The linear equations can be solved by refining its global solution
such that it is numerically stable. Both the wave-function and the
transmission spectra can be directly solved with high accuracy by our new method.
In addition to the new ability of calculating the \emph{wave-functions of higher dimensional
disordered systems}, our method is more efficient than the traditional TMM since the
stabilizing processes, which are most time-consuming~\cite{ROP1,ROP2,ROP3,MSF,MSF1} in the
traditional method, are not needed in our method.
In contrast to the diagonalization method, our method not only
provides a route for studying the wave-functions of \emph{open systems}
(corresponding to realistic transport experiments),
but also shows much higher efficiency to obtain the wave-function of a given energy/frequency.
In our method the calculating wave energy/frequency can be tuned continuously
while in the diagonalization method the wave-functions are fixed with discrete
eigenvalues. Those advantages make our method be very suitable to study the
mode-coupling effects in higher dimensional disordered systems.
As an example to show those advantages, we use it to identify the NS in a two dimensional(2D)
disordered Anderson model. This is the first report of observation of the NS
in the 2D Anderson model. Our method can be readily generalized to various
transport studies, such as electromagnetic/elastic waves propagation, or other models of
electronic systems~\cite{ChenTAI}.

This paper is arranged as follows. In Sec. II, we illustrate our new
formulization with the 2D Anderson model. (For models with spin-orbital
coupling, see our other work in Ref.~\cite{ChenTAI2011}.)
In Sec. III, we first compare the results of our method with the traditional TMM
and the direct diagonalization method. Then we discuss the memory usage and
computational efficiency of our method. Finally, we apply our method to
identify the NS in the 2D Anderson model. In Sec. IV, we conclude this paper.

\begin{figure}
\includegraphics[width=1.0 \columnwidth ]{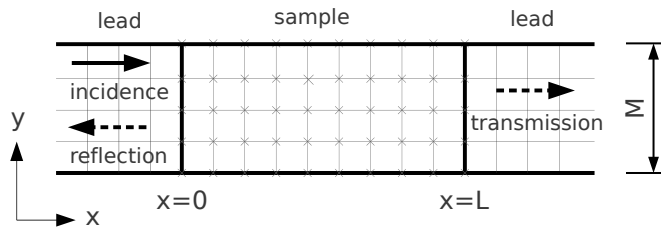}
\caption{\label{fig:epsart}
A quasi-1D sample attached by two semi-infinite clean leads on two sides.
The sample length is $L$ and the cross section is $M$ (in units of lattice constant).}
\end{figure}
\section{The Formulization}
To give a demonstration, we consider the 2D Anderson
tight binding model with diagonal disorder. Consider a quasi-1D sample with
length $L$ and cross section $M$(measured in units of lattice constant),
attached by two semi-infinite clean leads on two sides with the same width,
as schematically illustrated in Fig.1.
Cutting the strip into slices along the propagating direction $x$, the
traditional TMM commonly expresses the transport problem as a recursion~\cite{ROP1,ROP2,ROP3},
\begin{eqnarray}
\left(
\begin{array}{c}
\Psi(x+1)\\
\Psi(x)
\end{array}
\right) = \hat{M}(x) \left(
\begin{array}{c}
\Psi(x)\\
\Psi(x-1)
\end{array}
\right),
\label{TMM}
\end{eqnarray}
where $\Psi(x)=(\Psi_{x,1},...,\Psi_{x,y},\Psi_{x,M})$ is a vector of length
$M$, $\Psi_{x,y}$ represents the wave-function on site $(x,y)$,
and $\hat{M}(x)$ is the transfer matrix of slice $x$,
\begin{eqnarray}
\hat{M}(x)= \left(
\begin{array}{cc}
[E {\bf I}-\hat{H}(x)] & -{\bf I}\\
{\bf I} & {\bf O}\\
\end{array}
\right),
\label{TM}
\end{eqnarray}
with $E$ being the wave energy measured in units of the overlapping energy ($t$) between
neighboring sites(we set $t=1$). {\bf I} and {\bf O} are respectively unit and zero
matrices with dimension $M\times M$. $\hat{H}(x)$ is the ``slice Hamiltonian'' matrix:
\begin{eqnarray}
\hat{H}(x)= \left(
\begin{array}{ccccc}
\epsilon_{x,1} & 1 &  &  & 1\\
1 & \epsilon_{x,2} & 1 &  & \\
 & \cdot & \cdot & \cdot &  \\
 & & 1 & \epsilon_{x,M-1} & 1\\
1 & & & 1 & \epsilon_{x,M}\\
\end{array}
\right),
\end{eqnarray}
where the blanks denote zero, and the lower-left and the upper-right non-zero
elements reflect periodic boundary conditions along the $y$ direction.
$\epsilon_{x,y}$ is the on-site energy(also in units of the overlapping energy $t$),
which is set as zero in the clean leads and a random number uniformly
distributed in $[W/2, W/2]$ in the disordered sample. The boundary
conditions are described in the basis of eigenstates of clean
leads~\cite{ROP1,ROP2,ROP3,MSF}, $\hat{R}$, ( i.e., the eigenvectors of $\hat{M}_{lead}$)
\begin{eqnarray}
\left(
\begin{array}{c}
\hat{\Psi}(L)\\
\hat{\Psi}(L+1)
\end{array}
\right) = \hat{R}
\left(
\begin{array}{c}
{\bf O}\\
\hat{t}
\end{array}
\right) =
\left(
\begin{array}{cc}
\hat{R}_{11} & \hat{R}_{12}\\
\hat{R}_{21} & \hat{R}_{22}
\end{array}
\right)
\left(
\begin{array}{c}
{\bf O}\\
\hat{t}
\end{array}
\right),
\label{Boundary1}
\end{eqnarray}
\begin{eqnarray}
\left(
\begin{array}{c}
\hat{\Psi}(0)\\
\hat{\Psi}(1)
\end{array}
\right) =
\left(
\begin{array}{cc}
\hat{R}_{11} & \hat{R}_{12}\\
\hat{R}_{21} & \hat{R}_{22}
\end{array}
\right)
\left(
\begin{array}{c}
\hat{r}\\
\hat{I}
\end{array}
\right),
\label{Boundary2}
\end{eqnarray}
where the columns of $\hat{R}$, $\hat{R}(k)$'s, are ordered as
$k=1,2,\cdots,M$ correspond to modes propagating or decaying to
the left direction and $k=M+1,M+2,\cdots,2M$ correspond to
modes propagating or decaying to the right.
$\hat{R}_{ij}$'s $(i,j=1,2)$ are simply four $M\times M$ sub-blocks of $\hat{R}$.
Detailed descriptions of $\hat{R}$ have been given in many references, e.g.,
Refs.~\cite{ROP2,ROP3,MSF}. $\hat{t}$ and $\hat{r}$ respectively give the transmission
and reflection matrix with respect to incident waves carrying unit currents
(in each mode) from the left. The transmission coefficient is given by
$T=\sum_{i,j=1}^{N_0}|\hat{t}_{i,j}|^2$, where $N_0$ is the number of
propagating modes in the leads.
Since the transfer matrix is non-Hermitian, we need also the
left eigenvectors of $\hat{M}$, $\hat{L}$, the rows of which are
ordered the same as $\hat{R}$.
Usually the right half of $\hat{R}$ gives initial vectors for the iteration
in eq.(\ref{TMM}) and $\hat{t}^{-1}$ is projected out by the lower half
of $\hat{L}$ at the end of the iteration~\cite{ROP1,ROP2,ROP3,MSF}.

A numerical problem arises when carrying out the iterations in
eq.(\ref{TMM})~\cite{ROP1,ROP2,ROP3}. Since the eigenvalues of the disordered
transfer matrices are of the form $e^{\pm \alpha}$, on iterating
eq.(\ref{TMM}), the operating vectors rise exponentially with a set of
\emph{different} exponents $\alpha_i$'s. Those ones that rise the slowest, however,
and have the largest contributions to $\hat{t}$, will be inaccurate after $N$
iterations, provided that $e^{N(\alpha_{min}-\alpha_{max})}<\epsilon$,
where $\alpha_{max}(\alpha_{min})$ is the largest(smallest) exponent and
$\epsilon$ is the accuracy of the computer~\cite{ROP2}.
This can be effectively overcome by a
stabilizing process called ``re-orthogonalization''~\cite{ROP1,ROP2,ROP3},
in which the information of those vectors that rise slowest are reserved.
But then those vectors rising fastest, which are important for constructing
the wave-functions, are tend to be inaccurate. To calculate the wave-functions,
we try to get rid of the iteration form of eq.(\ref{TMM}) while preserving
the boundary conditions eq.(\ref{Boundary1},\ref{Boundary2}).
Note that $\hat{t}$ and $\hat{r}$ in eq.(\ref{Boundary1},\ref{Boundary2})
can be eliminated, giving
\begin{equation}
\hat{\Psi}(1)=\hat{R}_{21}\hat{R}_{11}^{-1}[\hat{\Psi}(0)-\hat{R}_{12}]+\hat{R}_{22},
\end{equation}
\begin{equation}
\hat{\Psi}(L)=\hat{R}_{12}\hat{R}_{22}^{-1}\hat{\Psi}(L+1),
\end{equation}
where $\hat{R}_{11}^{-1},\hat{R}_{22}^{-1}$ are the inverse matrices of
$\hat{R}_{11},\hat{R}_{22}$. Also note that the iteration of eq.(\ref{TMM})
is essentially from the discrete Schrodinger equation,
\begin{equation}
\hat{\Psi}(x-1) + (\hat{H}(x)-E)\hat{\Psi}(x) + \hat{\Psi}(x+1) = {\bf O}
\end{equation}
such that the whole description of the traditional TMM,
eq.(\ref{TMM},\ref{Boundary1},\ref{Boundary2}), can be embodied
in one set of linear equations self-consistently as follows,
\begin{widetext}
\begin{eqnarray}
\left(
\begin{array}{cccccc}
-\hat{R}_{21}\hat{R}_{11}^{-1} & {\bf I} & & & & \\
{\bf I} & \hat{H}(1)-E{\bf I} & {\bf I} & & & \\
 & {\bf I} & \hat{H}(2)-E{\bf I} & {\bf I} & & \\
 & & \cdot & \cdot & \cdot & \\
 & & & {\bf I} & \hat{H}(L)-E{\bf I} & {\bf I}\\
 & & & & {\bf I} & -\hat{R}_{12}\hat{R}_{22}^{-1 }
\end{array}
\right)
\left(
\begin{array}{c}
\hat{\Psi}(0)\\
\hat{\Psi}(1)\\
\hat{\Psi}(2)\\
 \cdot\\
\hat{\Psi}(L)\\
\hat{\Psi}(L+1)
\end{array}
\right) 
=
\left(
\begin{array}{c}
\hat{R}_{22}-\hat{R}_{21}\hat{R}_{11}^{-1}\hat{R}_{12}  \\
{\bf O}\\
{\bf O}\\
 \cdot\\
{\bf O}\\
{\bf O}
\end{array}
\right)\;,
 \label{LEqs}
\end{eqnarray}
\end{widetext}
where $\hat{\Psi}(i)$'s are the wave-functions needed to be solved.
This transformation is non-trivial since eq.(\ref{LEqs})
can be solved \emph{globally} by refinement of an initial guess of
the solution~\cite{NumericalRecipes} but not unidirectional substitution from one
end, where the former is numerically stable while the latter is similar
to the iteration of eq.(\ref{TMM}), being unstable.
With the solutions of $\hat{\Psi}(i)$'s, the transmission and
reflection matrices are then given by,
\begin{equation}
\hat{t}=\hat{R}_{22}^{-1}\hat{\Psi}(L+1),
\end{equation}
\begin{equation}
\hat{r}=\hat{R}_{11}^{-1}[\hat{\Psi}(0)-\hat{R}_{12}].
\end{equation}

\section{Applications}
\begin{figure}
\includegraphics[width=1.0 \columnwidth ]{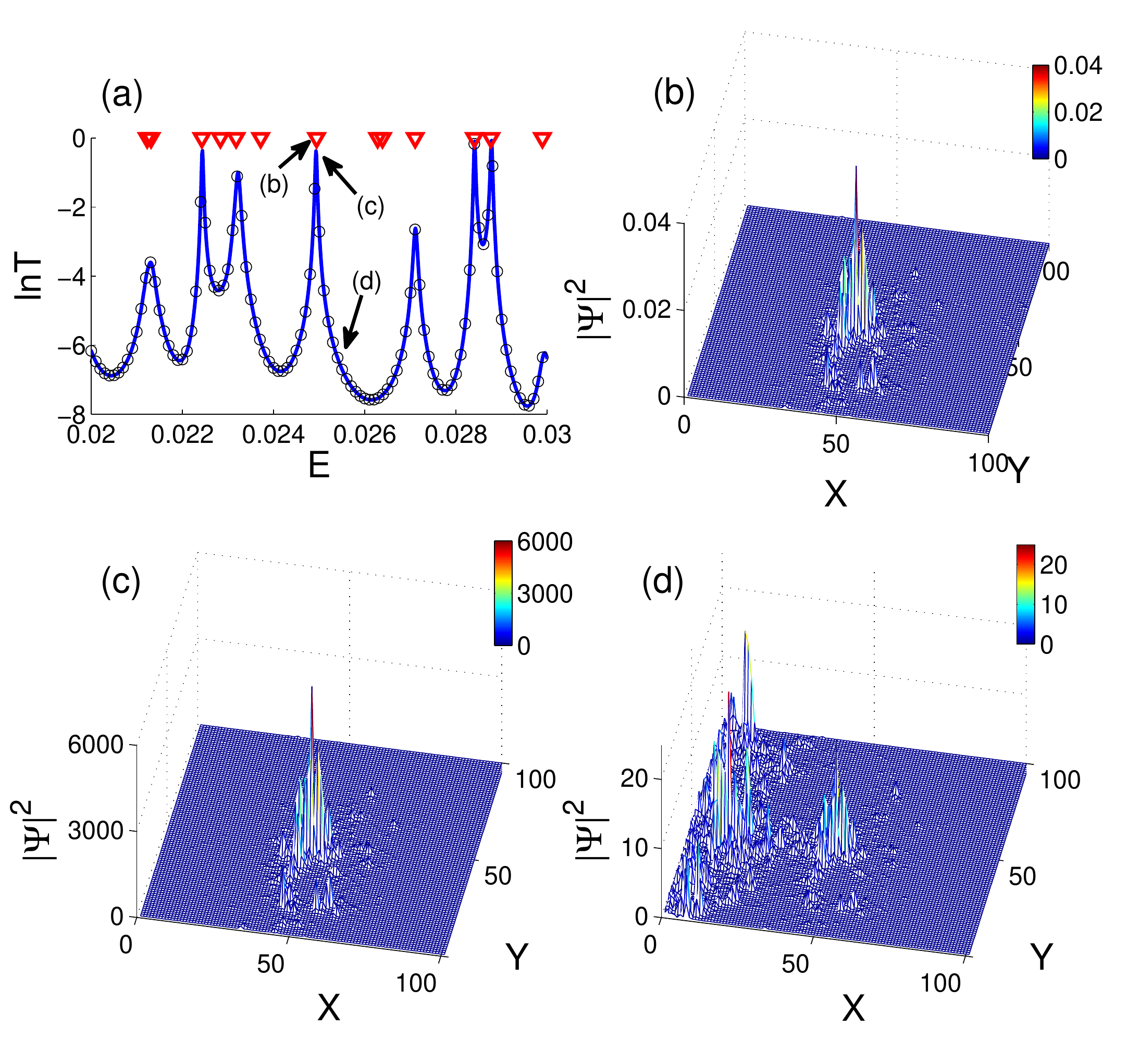}
\caption{(Color online )
(a): Transmission spectra calculated from our method (solid line) and the traditional
TMM (circles). Red triangles denote eigenvalues resulting from diagonalization,
where periodic boundary conditions are used in both $x$ and $y$ directions.
(b): $|\Psi|^2$ corresponding to the eigenenergy $E=0.02493$ calculated
from diagonalization. (c): $|\Psi|^2$ at the resonant energy (the same as
(b)) calculated from our method. (d): $|\Psi|^2$ at an off-resonant
energy ($E=0.0254$) calculated from our method.
We note that the wave-function in the traditional TMM diverges and
the diagonalization method cannot calculate the off-resonant case.}
\end{figure}
\subsection{Test Transmission Spectra and Wave-functions}
To test our method, we calculate a random configuration with size
$L\times M=100\times100$ and disorder strength $W=6$.
Fig.2(a) shows the transmission spectra calculated from our method (solid line)
and the traditional TMM (circles), where the ``re-orthogonalization''
is carried out in each iteration step in the traditional TMM.
We can see these two results match perfectly with each other.
In the 2D disordered Anderson model all eigenstates are localized.
Each transmission peak in Fig.2(a) represents a localized state and the valley
between peaks reflects the coupling strength between localized states
(and also the coupling condition to the environment)
~\cite{OpenSystem1,OpenSystem2,Pendry,Ghulinyan2007,Chen2011NJP}.
Those states strongly localized near the sample center are insensitive
to the boundary conditions in the sense that their eigenenergies and
wave-functions are almost unchanging when changing the boundary
conditions~\cite{Thouless1977}, such that we expect the transmission peaks
should roughly correspond to the eigenenergies obtained from diagonalizing
the same configuration with some artificial boundary conditions.
The eigenenergies obtained from diagonalizing the same configuration
(with periodic boundary conditions in both $x$ and $y$ directions)
are plotted by the red triangles in Fig.2(a).
Indeed, we can see each transmission peak precisely points to one
diagonalized eigenenergy. We note that the number of eigenenergies
is larger than the number of transmission peaks and some eigenenergies correspond to
no transmission peak. This is due to the differences in boundary conditions,
for example, some eigenstates cannot be excited from the $x$ direction but
may be easily excited in the $y$ direction.

More importantly, the wave-functions of our method are compared to the
eigenvectors of diagonalization. Because of the numerical instability,
the wave-functions in the traditional TMM diverge quickly (typically after
$10$ to $20$ iterations). Although the boundary conditions of the
diagonalization are much different from the TMM, for a state strongly
localized near the sample center, the diagonalized eigenvector still
provides us an effective reference substance~\cite{Thouless1977}.
Fig.2(b) shows the $|\Psi|^2$ distribution
corresponding to the eigenvector at eigenenergy $E=0.02493$. This
eigenenergy corresponds to a high transmission peak indicated by the
black arrow in Fig.2(a). By choosing $E$ the same as this eigenenergy,
the wave-function calculated from our method is shown in Fig.2(c),
where each mode is assumed to have an equal contribution to the total
wave-function. We can see these two wave-functions are in excellent agreement.
Since the localized state is very close to the sample center,
the resonant transport leads to a very high transmission peak.
Such a state is exactly the 2D Azbel state~\cite{Azbel1983}.
Moreover, since the energy in our method can be turned continuously,
we can also calculate the off-resonant wave-functions. Fig.2(d)
shows such a case, where $E=0.0254$ and $T=1.33\times 10^{-3}$.
It shows very clearly the localized state can only be partially excited
such that the wave-function amplitude decays significantly along the $x$ direction.
For those localized states far away from the sample center,
we observe a significant reduction of the peak value and also an
increasing of the peak width, which also agrees with the
understanding of resonant transport in 1D localized systems~\cite{Azbel1983}.
The consistency between our method and both the traditional TMM and
the diagonalization method proves that our method is confidently reliable.

\subsection{Memory Usage and Computational Efficiency}
Formally, the main part of the matrix on the left side of eq.(\ref{LEqs})
contains the Hamiltonian of the whole 2D system, while in the
traditional TMM we only need to store the Hamiltonian of several slices.
Hence the memory usage of our method is much larger than the traditional TMM
and is comparable to the diagonalization method.
The traditional TMM can in principle calculate infinitely long systems
while the system size in our method is limited.
However, the main matrix on the left side of eq.(\ref{LEqs}) is
very sparse such that one can still achieve very large systems if
sparse matrix technologies are used. The exact memory test will depend
on the algorithm implementation for solving sparse linear equations.
With the PARDISO solver~\cite{PARADISO}, we have achieved
a system size up to $M\times L=1600\times1600$ with 64GB computer memory for
a 2D system, which is good enough to perform most studies, except the finite size
scaling analysis of the localization length of quasi-1D systems, which can be
replaced by the scaling analysis of the Landauer conductance of a
hypercube~\cite{FSS,ChenTAI2011} (i.e., square/cube for two/three dimensions)
in our method.

\begin{figure}
\includegraphics[width=1.0 \columnwidth ]{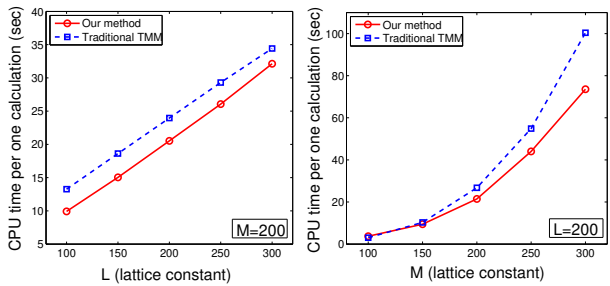}
\caption{\label{fig:epsart} (Color online ) Comparison of the
averaged CPU time between the traditional TMM and our new method. In the left figure
the system width $M$ is fixed at $200$ and in the right figure the system
length $L$ is fixed at $200$.}
\end{figure}
We next compare the efficiency of our new method with the traditional TMM.
We have implemented our method with the PARDISO solver~\cite{PARADISO} for
solving sparse linear equations and the traditional TMM with LAPACK
package~\cite{LAPACK} for inverting matrices. For the traditional TMM
the ``re-orthogonalization'' is performed after each $7$ iterations.
Fig.3 shows the averaged CPU time of both methods for one calculation(one energy)
on a 2.53GHz Intel(R) Xeon(R) processor.
In Fig.3(a) the system width $M$ is set as $200$ and $L$ is changed from $100$ to $500$.
The solid line marked by circles corresponds to our method and
the dashed line marked by squares corresponds to the traditional TMM.
It shows for both methods the CPU time scales almost linearly
with system length. Our method is moderately more efficient than the
traditional TMM. In Fig.3(b) the system length $L$ is set as $200$ and
$M$ is changed from $100$ to $300$. On increasing the system width $M$,
which determines the size of the matrices needed to be inverted in the
``re-orthogonalization'' process of the traditional TMM,
the advantage of our method becomes clear.
The main reason for this performance is that the matrix inversions of
the ``re-orthogonalization'', which will dominate the CPU time of the
traditional TMM~\cite{ROP1,ROP2,ROP3}, are not needed in our method.

Since solving linear equations is usually faster than diagonalizing matrices
of the same size, we do not discuss the diagonalization method here.
Actually, in many cases we only care about a very small energy region and
it is very inefficient to diagonalize the full Hamiltonian to get all
eigenenergies of the system. For example, using LAPACK to diagonalize
a $L\times M=100\times 100$ system(the matrix dimension is $10^4\times 10^4$),
one needs $5-6$ hours on a 2.53GHz Intel(R) Xeon(R) processor, while our
method only needs $2-3$ hours to calculate $10000$ arbitrary energy points.

To be brief, the numerical examinations show that our method is moderately
more efficient than the traditional TMM, especially for wider systems.
Our method is unique for calculating the wave-functions of higher dimensional
disordered open systems and is superior for treating finite systems. The
traditional TMM, on the other hand, has advantages in that it uses much less memory
and it can in principle calculate infinitely long systems. It is particularly useful
for applying the finite size scaling analysis of the localization length of
quasi-1D systems~\cite{FSS}.

\begin{figure}
\includegraphics[width=1.0 \columnwidth ]{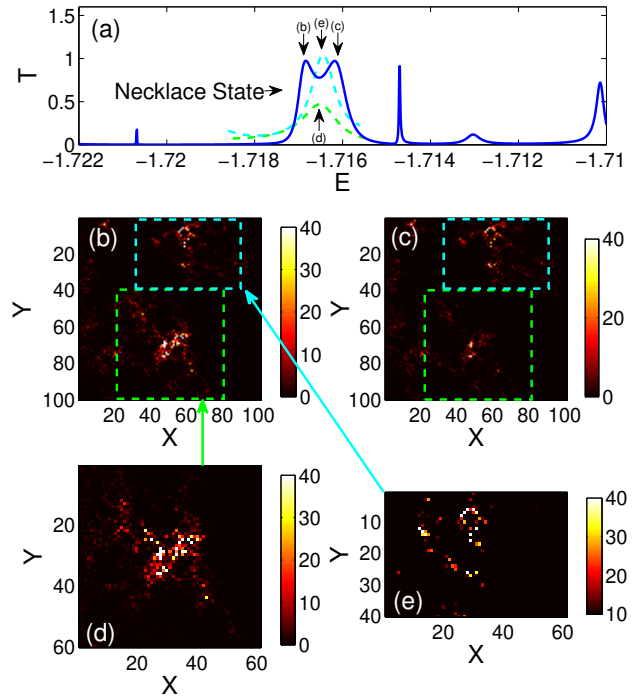}
\caption{\label{fig:epsart} (Color online ) A necklace state and its
decomposition. (a) Transmission spectrum of a $100\times100$ sample
with $W=6$(solid). The coupled peak represents a necklace state.
The dashed curves represent the resonant peaks of two localized states
shown in (d) and (e).
(b-c): $|\Psi|^2$ at the left and right sub-peaks of the necklace state shown
in (a). (d-e): $|\Psi|^2$ of two localized states in the two sub-samples
marked by the dashed rectangles in (b) (or (c)). The sub-sample sizes are $60\times60$
in (d) and $60\times40$ in (e). As shown by the dashed resonant peaks in (a),
the two localized states in (d) and (e) are nearly degenerate.}
\end{figure}
\subsection{Necklace state in the 2D Anderson model}
We next report an identification of the NS in the 2D Anderson model,
which also serves as an illustration of the powerfulness of our new method.
NSs~\cite{Pendry,Tartakovskii} are quasi-extended states formed from
degenerate coupling of localized states. They can contribute very high
transmission ``mini-bands'' and dominate the transmission of localized systems.
Their fundamental properties and statistical evidence are widely
studied in 1D systems~\cite{Pendry,Tartakovskii,Bertolotti2005PRL,Sebbah2006PRL,
Bertolotti2006PRE,Ghulinyan2007,Bliokh2008PRL,LiWei2009,Zhang2009PRB,Chen2011NJP}
and similar studies for higher dimensional systems are insistently
demanded~\cite{Pendry,Chen2011NJP,Mookerjee1993,ShengScience,PendryNature,Vanneste2009PRA}.
The 1D NS is characterized by the ``chain of localized states'' from the spatial
wave-functions and continuous phase evolution in its transmission ``mini-band''.
However, in 2D systems, the NS is hard to be characterized as continuous
phase evolution in the transmission spectrum as in the 1D
case~\cite{Bertolotti2005PRL,Ghulinyan2007,Chen2011NJP} since the phase
distribution at the outgoing interface is random. Here we identify the 2D NS by
making certain the degenerate coupling origin of the high transmission ``mini-band''.

A high transmission ``mini-band'' is shown in Fig.4(a), at $E=-1.7165$.
Here the ``mini-band'' exhibits as a ``coupled peak'' with two sub-peaks. The
electron densities $|\Psi|^2$ at those two sub-peak values, $E=-1.71683$ and
$E=-1.71617$, are shown in Fig.4(b) and Fig.4(c) respectively.
Fig.4(b) and (c) show that the $|\Psi|^2$ of each sub-peak has two localization
centers (the two high-brightness regions in each sub-figure).
The localization centers of the two sub-peaks are very similar,
showing characteristics of bonded and anti-bonded states formed from mode
coupling~\cite{Bliokh2008PRL}. To make certain the degenerate coupling's origin,
we take out two sub-samples around those two localization centers,
marked by the two dashed rectangles in Fig.4(b) (or (c)),
and calculate their transmission spectra. Interestingly, the transmission
spectra of both sub-samples show resonant peaks(green and cyan
dashed curves in Fig.4(a)) at the same energy $E=-1.7165$,
which are very close to the central energy of the coupled peak.
Electron densities of the corresponding localized states in the sub-samples
are shown in Fig.4(d) and Fig.4(e).
We can see those two localized states in the two sub-samples match up
to the two localization centers of Fig.4(b) and (c) very well.
Summing up those observations, if we combine the two sub-samples containing
degenerate localized states (d) and (e) into one larger sample, we observe a
coupled peak with two sub-peaks (b) and (c).
Such a coupled state is exactly the NS in the 2D Anderson model.
Similar to the 1D NS, it contributes a very large transmission ``mini-band''.
But the structure of the 2D NS seems more complicated than the 1D one. The
example in Fig.4 shows the positions of localized states could be
irregular, unlike the 1D case where the localized states are evenly
distributed along the sample.
We note that the implement as above can hardly be realized by the direct
diagonalization method. It is very hard to detect the mode coupling from
the discrete eigenenergies of closed systems.

\section{Summary and Discussions}
In summary, we have presented a new formulation of the TMM, which reformulates the traditional TMM with a set of
linear equations. Although the reformulation is very simple, the new method adequately gains many desired advantages.
Comparing with the traditional TMM, which is numerically unstable and can only calculate transmission of higher dimensional
disorder systems when applying some stabilizing processes (such as the ‘‘re-orthogonalization’’ process), our new method
is numerically stable such that it can accurately calculate both the transmission and the wave-functions. Without the need
of stabilizing processes, our method also shows advantages in that it is more efficient and simpler to implement. Comparing
with the direct diagonalization method, which also can calculate wave-functions of higher dimensional disordered systems, our
method not only provides a route for studying the wave functions of open systems corresponding to realistic transport experiments, but also shows much higher efficiency in calculating the wave-functions at arbitrary specific energies. That is to say, the calculating energy is controllable in our method, while in the diagonalization the eigenenergies are determined by the Hamiltonian and boundary conditions and one usually needs to solve all eigenenergies.

We have also used the new method to identify the NS in the two dimensional Anderson model. It shows the NS in a higher dimensional system can have much more complex structure than the 1D NS. The advantage of cooperating with both the wave-functions and the continuous transmission spectra makes our method very suitable for studying mode-coupling effects in higher dimensional disordered systems. It is straightforward to generalize our method to three dimensional or other modeling systems, such as models containing spin–orbit coupling~\cite{ChenTAI}, where interesting phenomena will appear in the presence of the Anderson transition~\cite{MirlinRMP}. The multifractality of wave-functions near the Anderson transition~\cite{MirlinRMP} can also be studied with our new method. It would be interesting to use our method to study the multifractal behavior of the wave-functions of open systems, which are connected with the wave-packets in transport experiments (usually the wave-packets can be constructed by linear combination of the diagonalized eigenstates)~\cite{Schreiber1993}, and compare them with those of the diagonalized eigenstates.

\section{Acknowledgments}
We would like to thank Dr. W. Li, Dr. Z. Liu and Dr. Q. Liu for helpful discussions.
This work is supported by the NSFC (Grant Nos. 11004212, 11174309, and 60938004),
and the STCSM (Grant Nos. 11ZR1443800 and 11JC1414500).



\begin{thebibliography} {s1}
\bibitem{MoreTMM} The basic technique is similar for other types of
differential equations. For more examples, see Ref.~\cite{TMMbook}.
\bibitem{MSF1} For electromagnetic waves, the numerical problem can be
alternatively solved by applying the ``multiple scattering formulism''~\cite{MSF}.
In this process the system is divided into successive sections(for details, see Ref.~\cite{MSF}),
such that one can only obtain correct wave-function in each short section(typically with
length of $10-20$ layers). Also, practical tests show that this process is
computationally very expensive, even more expensive than the ``re-orthogonalization process''.
\end{thebibliography}
\end{document}